\documentclass[aps,floatfix,showpacs,nofootinbib,showkeys,superscriptaddress,notitlepage,twocolumn]{revtex4}
\usepackage[T1]{fontenc}
\usepackage[english]{babel}
\usepackage[dvips]{graphicx}
\usepackage{amsmath,amssymb,amsfonts,mathtext,enumerate,float,mathrsfs}%символы всякие
\usepackage{wrapfig}
\usepackage{hhline}
\usepackage{epstopdf}

\usepackage{epsfig}
\usepackage{epsf}
\usepackage{subfig} % for figures in a number
\usepackage{feynmp} % for to create diagram Feyman
\usepackage{pgf,pgfarrows,pgfnodes,pgfautomata,pgfheaps,multirow}
\usepackage{array}
\usepackage{bm}% bold math \
\usepackage{latexsym}
\usepackage{pifont}

\begin{document}
\title{Radially Excited Axial-Vector Mesons in the extended Nambu--Jona-Lasinio model}

\author{A.\ V.\ Vishneva}
\email{vishneva@theor.jinr.ru}
\affiliation{Bogoliubov Laboratory of Theoretical Physics,
JINR, Dubna, 141980  Russia}

\author{M.\ K.\ Volkov}
\email{volkov@theor.jinr.ru}
\affiliation{Bogoliubov Laboratory of Theoretical Physics,
JINR, Dubna, 141980  Russia}

\begin{abstract}

The first radial excitations of axial-vector mesons are considered in the framework of the extended $U(3)\times U(3)$ Nambu--Jona-Lasinio model. We calculate the mass spectrum of $a_1$, $f_1$, and also strange axial-vector mesons. For description of radially excited states we used the form factors of polynomial type of the second order in transverse quark momentum. For the ground- and the excited-state mesons consisting of light quarks we have calculated the widths of a number of strong and radiative decays. We got satisfactory agreement with experimental data for the ground states. A set of predictions for the excited states of mesons is given.  
\end{abstract}

\keywords{axial-vector mesons, radially excited mesons, Nambu--Jona-Lasinio model, radiative decays, strong decays} 

\pacs{12.39.Fe, 13.20.Jf, 13.25.Jx}

%\begin{minipage}{\linewidth}
\maketitle
%\end{minipage}

\section{INTRODUCTION}

The extended Nambu--Jona-Lasonio (NJL) model is well known to describe scalar, pseudoscalar and vector mesons in the ground and first radially excited states \cite{weiss,yaf,ven,yud,ufn}. In recent papers, the processes of meson production in the $e^+ e^-$ annihilation at energies below 2 GeV were successfully studied in the framework of this model (see \cite{last} and references therein). In these works, intermediate vector mesons in the ground and first radially excited states were taken into account.

The same mechanism is also used for describing the $\tau$-lepton decays \cite{last,rhoeta,omega,taupipi}.
However, unlike the $e^+ e^-$ processes, there is a number of $\tau$-decays with intermediate axial-vector mesons, for instance, $\tau\rightarrow 3\pi\nu_\tau$ \cite{zph,Dumm:2009va}, $\tau\rightarrow \pi\gamma\nu_\tau$ \cite{taupig,Guo:2010dv}, $\tau\rightarrow \pi l^+ l^- \nu_\tau$ \cite{ll} $\tau\rightarrow f_1 \pi\nu_\tau$ \cite{Li:1996md,tauap}.

In all these works only the ground states of axial-vector mesons were taken into account. In order to consider also radially excited states of these mesons, we need to include axial-vector mesons in the extended NJL model. This paper is devoted to solving this problem.

Here we consider the chiral $U(3)\times U(3)$ group including mesons consisting of u, d, and s-quarks. Radial excitations are described with the polynomial form factor of the second order in transverse quark momentum. 

The article is organized as follows. In the next section, we obtain the physical quark-meson Lagrangian and expressions for physical meson fields and masses. 
In Section III, we test our Lagrangian by considering various decays involving $a_1$ and $f_1$ mesons, and also vector, scalar, and pseudoscalar mesons in the ground and the first radially-excited states.
In Conclusion, we briefly discuss our results and possibilities for their further application. In Appendix, we give a list of coefficients used in this work.

\section{The Effective Lagrangian for axial-vector mesons in the ground and the excited states}
\label{L}

First we consider $SU(2)\times SU(2)$ extended NJL model to show how to get the physical quark-meson Lagrangian and the meson masses. The part of the four-quark NJL Lagrangian corresponding to vector and axial-vector mesons reads \cite{86,weiss,yaf,ufn}
\begin{eqnarray}
\label{eq:4q}
 \Delta\mathcal{L} (\bar{q},q)&=& -\frac{G_V}{2} \left[j^{\mu,a}_{v_1}(x) j^{\mu,a}_{v_1}(x)+j^{\mu,a}_{v_2}(x) j^{\mu,a}_{v_2}(x)\right] \nonumber\\
 &-&\frac{G_A}{2}\left[j^{\mu,a}_{a_1}(x) j^{\mu,a}_{a_1}(x)+j^{\mu,a}_{a_2}(x) j^{\mu,a}_{a_2}(x) \right],
\end{eqnarray}
where $G_V$ and $G_A$ are the four-quark coupling constants for vector and axial-vector mesons, $j^v_i$ and $j^a_i$ are vector and axial-vector quark currents defined as
\begin{eqnarray}
j^{\mu,a}_{v_1}(x)&=&\bar{q}(x)\gamma^\mu \tau^a q(x), \\
\label{eq:f}
j^{\mu,a}_{v_2}(x)&=&\int d^4 x_1 \int d^4 x_2 \bar{q}(x_1) \nonumber \\
&\times &\gamma^\mu \tau^a  F(x;x_1,x_2) q(x_2),\\
j^{\mu,a}_{a_1}(x)&=&\bar{q}(x)\gamma^\mu \gamma^5 \tau^a q(x), \\
j^{\mu,a}_{a_2}(x)&=&\int d^4 x_1 \int d^4 x_2 \bar{q}(x_1)\nonumber \\
&\times & \gamma^\mu \gamma^5 \tau^a  F(x;x_1,x_2) q(x_2),
\end{eqnarray}
where $q=\{u, d\}$ are quark fields, $\tau_{i=1,2,3}$ are the Pauli matrices and $\tau_0=\textbf{I}$. In Eq. \eqref{eq:f}, we introduced the form factor $F(x;x_1,x_2)$ which has a complicated form in the coordinate frame \cite{weiss}. However, in our calculations it will be used only in the momentum frame, where it reads $F(k_\bot)=c_a(1+d_a k_\bot^2)=c_a f(k_\bot)$, where $k_\bot=k-\frac{(kp)p}{p^2}$, $k$ and $p$ are the quark and meson momenta, respectively. In the rest frame of mesons $k_\bot=\textbf{k}.$ Here $c_a$ is a free parameter, and it influences meson masses. One can see that meson interaction does not require $c_a$. The parameter $d_a$ is -1.784 $\mathrm{GeV}^{-2}$ for vector and axial-vector mesons consisting of light quarks. It is defined from the condition that the excited states should not contribute to the vacuum expectation values of the quark fields (see \cite{weiss,yaf}). 

After the bosonization of the Lagrangian \eqref{eq:4q}, the quark-meson Lagrangian takes the form (in the momentum space)

\begin{eqnarray}
\Delta\mathcal{L} (\bar{q},q,\breve{v},\tilde{v},\breve{a},\tilde{a})&=&\frac{1}{2 G_V}((\breve{v}^a_\mu)^2+(\breve{a}^a_\mu)^2)\\ \nonumber
&+&\frac{1}{2 c^2_a G_V}((\tilde{v}^a_\mu)^2+(\tilde{a}^a_\mu)^2)\\ \nonumber
&+&\bar{q}(k') \gamma^\mu \tau_a \left(\breve{v}_a^\mu+f(k_\bot) \tilde{v} _a^\mu \right)q(k)
\\ \nonumber
&+&\bar{q}(k) \gamma^\mu \gamma^5 \tau_a \left( \breve{a}_a^\mu+ f(k_\bot) \tilde{a} _a^\mu \right)q(k),
\end{eqnarray}
where $\breve{v}$ ($\tilde{v}$) and $\breve{a}$ ($\tilde{a}$) are the ground-state (excited-state) vector and axial-vector meson fields, respectively. Namely, $v_{i=1,2,3}$ are $\rho$-mesons, $v_0$ is $\omega$-meson, $a_{i=1,2,3}$ are $a_1$-mesons, and $a_0$ denotes $f_1$. These fields are not physical yet, because they lead to non-diagonal terms in the free meson Lagrangian.
\begin{figure}[!htb]
       \centering
       \includegraphics[width=0.9\linewidth]{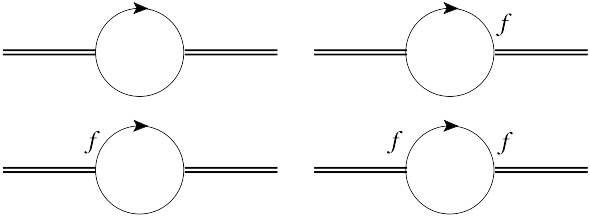}
       \caption{The quark loop contribution to the meson Lagrangian \eqref{eq:free}. Letter \textit{f} in the vertices denotes radially excited meson field.}
       \label{loops}
     \end{figure}
     
Indeed, after taking into account loop diagrams (see Fig. \ref{loops}) and renormalization of the meson fields the free Lagrangian for the vector and the axial-vector mesons takes the form \cite{yaf}
\begin{equation}
\label{eq:free}
\Delta\mathcal{L} (\breve{v},\tilde{v},\breve{a},\tilde{a})=\mathcal{M}_V+\mathcal{K}_V+\mathcal{M}_A+\mathcal{K}_A,
\end{equation}
where
$$
\begin{aligned}
\mathcal{M}_V&=\frac{M_1^2}{2}(\breve{v}^a_\mu)^2+\frac{M_2^2}{2}(\tilde{v}^a_\mu)^2,\\ 
\mathcal{K}_V&=-\frac12 \left(p^2 g^{\mu\nu}-p^\mu  p^\nu\right) \left(\breve{v}^a_\mu\breve{v}^a_\nu+2\Gamma\breve{v}^a_\mu \tilde{v}^a_\nu+\tilde{v}^a_\mu\tilde{v}^a_\nu\right),\\
\mathcal{M}_A&=\frac{M_1^2}{2}(\breve{a}^a_\mu)^2+\frac{M_2^2}{2}(\tilde{a}^a_\mu)^2,\\
\mathcal{K}_A&=-\frac12 \left(p^2 g^{\mu\nu}-p^\mu  p^\nu-6 m_u^2 g^{\mu\nu}\right)\\
&\times  \left(\breve{a}^a_\mu\breve{a}^a_\nu+2\Gamma\breve{a}^a_\mu \hat{a}^a_\nu+\tilde{a}^a_\mu\tilde{a}^a_\nu\right),\\ 
M_1^2&=\frac{g_\rho^2}{4 G_V},\quad M_2^2=\frac{g_{\rho'}^2 }{4 {c_a}^2 G_V}, 
\end{aligned}
$$
$g_\rho$ and $g_{\rho'}$ are the quark-meson coupling constants, which read

\begin{equation}
g_{\rho} = \left(\frac{2}{3} I_2(m_u)\right)^{-1/2},\quad g_{\rho'} = \left(\frac{2}{3} I_2^{f^2}(m_u)\right)^{-1/2},
\end{equation}
the integrals $I^{f^{n}}_m(m_q)$ have the form
\begin{equation}
\label{eq:i}
I^{f^{n}}_m(m_q) = \int\frac{\mbox{d}^4 k}{(2\pi)^4}\frac{(f_q({k^\bot}^2))^n}{(m_q^2-k^2)^m}\Theta(\Lambda^2_3 - \vec k^2),
\end{equation}
the cut-off parameter $\Lambda_3 = 1.03~$ GeV, $m_q = m_u=m_d=280~\mathrm{MeV}$ are the quark masses, and
$$\Gamma=\frac{I^f_2}{\sqrt{I_2 I^{f^2}_2}}.$$

Here we have used the following relations (see \cite{86,ufn}):
\begin{eqnarray}
\frac{\mathrm{Tr} \gamma^\mu\gamma^5 (m_u+\hat{k})\gamma^\nu\gamma^5(m_u+\hat{k}+\hat{p})}{(m_u^2-k^2)(m_u^2-(k+p)^2)}=\\ \nonumber
\frac{\mathrm{Tr} \gamma^\mu (m_u+\hat{k})\gamma^\nu(m_u+\hat{k}+\hat{p})}{(m_u^2-k^2)(m_u^2-(k+p)^2)}-6 m_u^2 g^{\mu\nu};\\ 
\frac{\mathrm{Tr} \gamma^\mu (m_u+\hat{k})\gamma^\nu(m_u+\hat{k}+\hat{p})}{(m_u^2-k^2)(m_u^2-(k+p)^2)}=\\ 
4 I_2 (p^2 g^{\mu\nu}-p^m p^\nu).\nonumber
\end{eqnarray}

One can see that the non-diagonal terms for vector and axial-vector mesons are similar.
In \cite{yaf} the diagonalized Lagrangian with physical vector meson states was obtained. In order to obtain the corresponding Lagrangian for the axial-vector mesons, which reads

\begin{eqnarray}
\Delta\mathcal{L} (a,a')&=&-\frac12 \left[\left(p^2 g^{\mu\nu}-p^\mu  p^\nu-M_a^2\right)a^a_\mu a^a_\nu\right.\\ \nonumber
&+&\left.\left(p^2 g^{\mu\nu}-p^\mu  p^\nu-M_{a'}^2\right){a'}^a_\mu {a'}^a_\nu\right],
\end{eqnarray}
we apply the same method. Namely, we transform the meson fields $\breve{a}$ and $\tilde{a}$ \cite{yaf,ven}:

\begin{eqnarray}
a = \cos(\chi-\chi_0)\breve{a}-\cos(\chi+\chi_0)\tilde{a},\\
a' = \sin(\chi-\chi_0)\breve{a}-\sin(\chi+\chi_0)\tilde{a}.
\end{eqnarray}
After this, one can get the expressions for the physical meson masses:

\begin{eqnarray}
M_a^2 &=& \frac{1}{2\sqrt{1-\Gamma^2}} \Big[M_1^2+M_2^2\\ \nonumber
&-&\sqrt{(M_1^2-M_2^2)^2+(2\Gamma M_1 M_2)^2} \Big]+6 m_u^2,\\
M_{a'}^2 &=& \frac{1}{2\sqrt{1-\Gamma^2}} \Big[M_1^2+M_2^2\\ \nonumber
&+&\sqrt{(M_1^2-M_2^2)^2+(2\Gamma M_1 M_2)^2} \Big]+6 m_u^2,
\end{eqnarray}
and the mixing angles:

\begin{eqnarray}
\label{eq:c0}
\sin \chi_0 &=& \sqrt{\frac{1+\Gamma^2}{2}},\\
\label{eq:c}
\tan(2\chi-\pi)& = &\sqrt{\frac{1}{\Gamma^2}-1}\left[\frac{M_{\breve{a}}^2-M_{\tilde{a}}^2}{M_{\breve{a}}^2+M_{\tilde{a}}^2}\right].
\end{eqnarray}

In this paper, we use $c=1.26$, the four-quark coupling constant $G_V=12.5~\mathrm{GeV}^{-2}$, and the mixing angles $\chi_0=~61.44^{\circ}$, $\chi=~79.85^{\circ}$. The same parameters were used for the vector mesons. Therefore, one can obtain values for the physical masses of $a_1$ and $a'_1$:

\begin{eqnarray}
M_{a_1}&=&1028~\mathrm{MeV},\\ \nonumber
M_{a'_1}&=&1647~\mathrm{MeV}.
\end{eqnarray}
One can see that $M_{a_1}$ in this model differs from its experimental value $M_{a_1(1260)}=1230\pm 40~\mathrm{MeV}$ noticeably. However, the first radial excitation of $a_1$ meson is described in good agreement with experimental data: $M_{a_1(1640)}=1647\pm 22~\mathrm{MeV}$.

Let us note that a similar situation takes place in description of isovector scalar mesons \cite{yud}. Indeed, the mass of the ground and excited states obtained theoretically are 830 MeV and 1500 MeV, respectively, while experimental values are $980\pm 20$ MeV and $1474\pm19$ MeV \cite{pdg}.

The same theoretical values correspond isoscalar axial-vector mesons $f_1$ and $f'_1$, while experimental values are $1281\pm 0.5$ MeV and $1518\pm 5$ MeV.

Finally, the $SU(2)\times SU(2)$ NJL Lagrangian describing interaction of the physical vector and axial-vector mesons with quarks takes the form:

\begin{eqnarray}
\label{eq:lva}
\Delta\mathcal{L}^{int}_{v,a} &=&\bar{q}(k')\tau^a \gamma^\mu \left[ A_1^\rho (v_a^\mu+a_a^\mu\gamma^5)\right.\\ \nonumber
&-&\left. A_2^\rho({v'}_a^\mu+{ a'}_a^\mu\gamma^5)\right]q(k),
\end{eqnarray}
where
\begin{eqnarray} 
A_1^\rho &=& g_\rho \frac{\sin(\chi+\chi_0)}{\sin(2\chi_0)}+g_{\rho'} f(k_\bot^2)\frac{\sin(\chi-\chi_0)}{\sin(2\chi_0)},\\ \nonumber
A_2^\rho &=& g_\rho \frac{\cos(\chi+\chi_0)}{\sin(2\chi_0)}+g_{\rho'} f(k_\bot^2)\frac{\cos(\chi-\chi_0)}{\sin(2\chi_0)}.
\end{eqnarray}

Let us now consider the $U(3)\times U(3)$ NJL model including also s-quarks. In this case, one has to use The Gell-Mann matrices instead of the Pauli matrices, namely, $\lambda_{i=1,7},~\lambda_u=\mathrm{diag}(1,1,0),~lambda_s=\mathrm{diag}(0,0,-\sqrt{2})$, where $\lambda_{i=1,2,3}$ denote $\tau_{1,2,3}$. Using the method considered in Section \ref{L}, one can obtain the following mass formulae for strange mesons
\begin{eqnarray}
M_{K_1}^2 &=& \frac{1}{2\sqrt{1-\Gamma^2}} \Big[(M_1^2+M_2^2\\ \nonumber
&-&\sqrt{(M_1^2-M_2^2)^2+(2\Gamma M_1 M_2)^2} \Big]+6 m_u m_s,\\
M_{K'_1}^2 &=& \frac{1}{2\sqrt{1-\Gamma^2}} \Big[M_1^2+M_2^2\\ \nonumber
&+&\sqrt{(M_1^2-M_2^2)^2+(2\Gamma M_1 M_2)^2} \Big]+6 m_u m_s,
\end{eqnarray}
and $\bar{s}s$ isoscalar:
\begin{eqnarray}
M_{\phi}^2 &=& \frac{1}{2\sqrt{1-\Gamma^2}} \Big[M_1^2+M_2^2\\ \nonumber
&-&\sqrt{(M_1^2-M_2^2)^2+(2\Gamma M_1 M_2)^2} \Big]+6 m_s^2,\\
M_{\phi'}^2 &=& \frac{1}{2\sqrt{1-\Gamma^2}} \Big[M_1^2+M_2^2\\ \nonumber
&+&\sqrt{(M_1^2-M_2^2)^2+(2\Gamma M_1 M_2)^2} \Big]+6 m_s^2.
\end{eqnarray}
The corresponding non-physical masses are
\begin{equation}
\left(M^{us,ss}_1\right)^2=\frac{g_{K^*,\phi}^2}{4 G_V},\quad \left(M^{us,ss}_2\right)^2=\frac{g_{{K^*}',\phi'}^2 }{4 {c_{K^*,\phi}}^2 G_V},
\end{equation}
where $c_{K^*}=1.5,~c_{\phi}=1.48$,
\begin{eqnarray}
g_{K^*} &=& \left(\frac{2}{3} I_2(m_u,m_s)\right)^{-1/2},\quad g_{\phi} = \left(\frac{2}{3} I_2(m_s)\right)^{-1/2}, \\ \nonumber
g_{{K^*}'} &=& \left(\frac{2}{3} I_2^{f^2}(m_u,m_s)\right)^{-1/2}, \quad g_{\phi'} = \left(\frac{2}{3} I_2^{f^2}(m_s)\right)^{-1/2},
\end{eqnarray}
the integral $I_2(m_s)$ is obtained from \eqref{eq:i}, and

\begin{equation}
I^{f^{n}}_2(m_u,m_s) = \int\frac{\mbox{d}^4 k}{(2\pi)^4}\frac{(f_q({k^\bot}^2))^n}{(m_u^2-k^2)(m_s^2-k^2)}\Theta(\Lambda^2_3 - \vec k^2).
\end{equation}
The corresponding slope parameters are $d_{us}=-1.756~\mathrm{GeV}^{-2}$ and $d_{ss}=-1.727~\mathrm{GeV}^{-2}$. One can also obtain the mixing angles using the formulae \eqref{eq:c0},~\eqref{eq:c}.

As a result, we have calculated the masses of strange axial-vector mesons. For $K_1$ mesons we obtain $M_{K_1}=1245~\mathrm{MeV}$ and $M_{K'_1}=1693~\mathrm{MeV}$, while the experimental values are $M^{exp}_{K_1}=1272\pm7~\mathrm{MeV}$ and $M^{exp}_{K'_1}=1650\pm50~\mathrm{MeV}$. For $\bar{s}s$ isoscalar meson we get $M_{f_1}=1510~\mathrm{MeV}$ and $M_{f'_1}=2017~\mathrm{MeV}$, and its ground-state mass is $M^{exp}_{f_1}=1426.4\pm0.9~\mathrm{MeV}$ \cite{pdg}. One can see that the results for strange mesons are better in comparison with light-quark mesons.

\section{Strong and radiative decays involving axial-vector mesons}

In this section, we use the Lagrangian \eqref{eq:lva} for describing of strong and radiative meson decays. For this, widths of a number of decays involving the ground- and excited-state axial-vector mesons are calculated. In order to describe the following processes, one needs an additional interaction Lagrangian with electromagnetic and scalar and pseudoscalar meson fields of the following form \cite{ufn}:

%full lagrangian
\begin{eqnarray}
\label{eq:lsp}
\Delta\mathcal{L}^{int}_{\gamma,\sigma, \pi}&=&\bar{q}(k')\Big[-e Q \hat{A}+\tau_i (A_1^{\sigma} \sigma_i-A_2^{\sigma} \sigma'_i)\\ \nonumber
&+& i \tau_i \gamma^5 (A_1^{\pi} \pi_i-A_2^{\pi} \pi'_i)\Big]q(k),\\ \nonumber
\end{eqnarray}
where $\sigma$ and $\pi$ are scalar and pseudoscalar meson fields, namely, $\sigma_{i=1,2,3}$ denote $a_0$, $\pi_i$ denote $\pi$-mesons, $e$ is the electron charge, $A$ is the electromagnetic field, $Q$ is the quark charge operator:
\begin{equation}
Q=\frac12 \left( \tau^3+\frac{\tau_0}{3}  \right),
\end{equation}

\begin{eqnarray}
A_1^{\sigma,\pi} &=& g_{\sigma,\pi} \frac{\sin(\alpha+\alpha_0)}{\sin(2\alpha_0)}+g_{\sigma',\pi'} f(k_\bot^2)\frac{\sin(\alpha-\alpha_0)}{\sin(2\alpha_0)},\\ \nonumber
A_2^{\sigma,\pi} &=& g_{\sigma,\pi} \frac{\cos(\alpha+\alpha_0)}{\sin(2\alpha_0)}+g_{\sigma',\pi'} f(k_\bot^2)\frac{\cos(\alpha-\alpha_0)}{\sin(2\alpha_0)},
\end{eqnarray}
where coupling constants
\begin{eqnarray}
g_{\sigma} &=& \left(4 I_2(m_u)\right)^{-1/2},\quad g_{\pi}=\sqrt{Z} g_{\sigma}\\ \nonumber
g_{\sigma'}&=&g_{\pi'} = \left(4 I_2^{f^2}(m_u)\right)^{-1/2},
\end{eqnarray}
and $Z=(1-6m_u^2/m_{a_1}^2)^{-1}$ is the factor corresponding to the $\pi-a_1$ transitions.

%a1rp figure
\begin{figure}[!htb]
       \centering
       \includegraphics[width=0.9\linewidth]{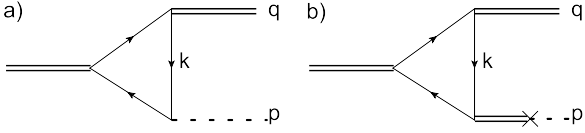}
       \caption{Diagram describing $a_1\rightarrow\pi\rho$ decay.}
       \label{decay}
     \end{figure}

At first, let us consider the decay $a_1\rightarrow \pi \rho$, described by the diagram on Fig. \ref{decay}, where $p$ and $q$ are the pion and $\rho$ meson momenta, respectively (the diagram 2b includes $\pi-a_1$ transition). In the framework of the standard NJL model, the amplitude of this process was obtained in \cite{arp}.
It has the form 

\begin{eqnarray}
\label{eq:a1pirho}
T_{a_1\rightarrow \pi \rho}^{S}&=&i\frac{\alpha_\rho}{2 \pi F_\pi}\Bigg[p^\mu q^\nu - g^{\mu\nu}(p q)\\ \nonumber
&+&g^{\mu\nu}\left(1+2Z\left(\frac{2 \pi F_\pi}{m_{a_1}}\right)^2\right)q^2\Bigg]{\epsilon^\mu_\rho}^* \epsilon^\nu_{a_1},
\end{eqnarray}
where $\alpha_\rho=g_\rho^2/4\pi$ and $F_\pi=93~\mathrm{MeV}$ is pion week decay constant.
%The decay width obtained in the framework of the standard NJL model is 203 MeV.

In the extended NJL model, the amplitude~\eqref{eq:a1pirho} becomes more complicated:

\begin{eqnarray}
T_{a_1\rightarrow \pi \rho}^{E}&=&\frac{i}{8 \pi^2 F_\pi}\Bigg[\left(p^\mu q^\nu - g^{\mu\nu}\right)\frac{V_4^{a_1 \pi\rho}}{I_4(m_u)}(pq)\\ \nonumber
&+&g^{\mu\nu}\left(M^2_\pi+M^2_\rho \right)\frac{V_3^{a_1 \pi\rho}}{I_3(m_u)}\\ \nonumber
&+& 6\pi^2 F^2_\pi g^{\mu\nu}\Bigg[\frac{V_2^{a_1 \pi\rho}}{I_2(m_u)}\\ \nonumber
&-&\left(1-\frac{M^2_\rho}{M^2_{a_1}}\right)\frac{V_2^{a_1 a_1 \rho}}{g_\rho I_2(m_u)}\frac{V_2^{a_1\pi}}{g_\rho I_2(m_u)} \Bigg]\Bigg]{\epsilon^\mu_\rho}^* \epsilon^\nu_{a_1},
\end{eqnarray}
where the coefficients $V$ are defined in Appendix. 
One can also obtain the amplitude of the decay $a'_1\rightarrow \rho \pi$ and some others, by simple replacing of the the corresponding meson masses and coefficients in this amplitude. The decay widths for a set of similar processes are given in Table I.

The next process we considered is $a_1\rightarrow\pi\gamma$. Its amplitude can be obtained from the amplitude \eqref{eq:a1pirho} by defining $q^2=0$ and in the standard NJL model it has the form (see \cite{arp})
\begin{equation}
T_{a_1\rightarrow \pi \gamma}^S=i \frac{e g_\rho}{8\pi^2 F_\pi} \left(p^\mu q^\nu - g^{\mu\nu}(pq)\right){\epsilon^\mu_\gamma}^* \epsilon^\nu_{a_1}.
\end{equation}
The corresponding amplitude in the extended NJL model reads
\begin{equation}
T_{a_1\rightarrow \pi \gamma}^E=i \frac{e V_4^{a_1\pi\gamma}}{4 \pi^2 m_u} \left(p^\mu q^\nu - g^{\mu\nu}(pq)\right){\epsilon^\mu_\gamma}^* \epsilon^\nu_{a_1}.
\end{equation}
We have got good agreement with experimental data for this process (see Table I).

Considering $f_1$ meson, we have primarily studied the decay $f_1\rightarrow a_0 \pi$, which is rather simple and widely explored experimentally. Its amplitude in the standard and the extended NJL model reads

\begin{eqnarray}
T_{f_1\rightarrow a_0 \pi}^S &=& g_\rho (p-q)_\mu \epsilon^\mu_{f_1},\\
T_{f_1\rightarrow a_0 \pi}^E &=&\frac{2}{\sqrt{6}} g_\rho V_2^{f_1 a_0 \pi} (p-q)_\mu \epsilon^\mu_{f_1}.
\end{eqnarray}
The decay width we calculated is also in good agreement with the experimental value (see Table I).

Finally, we have considered the decay $a'_1\rightarrow f_1\pi$, which is useful for the description of the decay $\tau\rightarrow f_1\pi \nu_\tau$. Its amplitude has the following form: 
\begin{equation}
T_{a'_1\rightarrow f_1\pi}^E=i\frac{3 V_3^{a'_1 \pi f_1}}{8 \pi^2 F_\pi}\epsilon_{\mu\alpha\nu\beta}p^\alpha q^\beta.
\end{equation}
The decay width of the process $f'_1\rightarrow a_1\pi$ is calculated in the same way.

\begin{table}
\caption{Decay widths}
\begin{tabular}{|c|c|c|}
\hline 
Decay & Extended NJL, MeV & PDG, MeV\cite{pdg} \\ 
\hline 
$a_1(1260)\rightarrow\rho(770)\pi$ & 148 & --- \\ 
\hline 
$a_1(1640)\rightarrow\rho(770)\pi$ & 8.5 & --- \\ 
\hline 
$\rho(1450)\rightarrow a_1(1260)\pi$ & 6.5 & --- \\ 
\hline 
$a_1(1260)\rightarrow\pi\gamma$ & 0.48 &  $0.64 \pm 0.246$ \\ 
\hline 
$a_1(1640)\rightarrow\pi\gamma$ & 0.04 & --- \\ 
\hline 
$a_1(1640)\rightarrow\pi(1300)\gamma$ & $1.2\times 10^{-3}$ & --- \\ 
\hline 
$\pi(1300)\rightarrow a_1(1260)\gamma$ & $0.31\times 10^{-3}$ & --- \\ 
\hline 
$f_1(1285)\rightarrow a_0(980) \pi$ & 11.4 & $8.7\pm 2.05$ \\ 
\hline 
$f_1(1510)\rightarrow a_0(980) \pi$ & 4.2 & --- \\ 
\hline 
$a_1(1640)\rightarrow f_1(1285)\pi$ & 0.16 & --- \\ 
\hline 
$f_1(1510)\rightarrow a_1(1260)\pi$ & 0.14 & --- \\ 
\hline 
\end{tabular} 
\end{table}

\section{Conclusion}

In this paper, we give the generalization of the extended NJL model by introducing the axial-vector mesons. %Here two opportunities appear. The first opportunity is considering parameters $G_A$ and $c_A$ as arbitrary ones for axial-vector mesons for a correct mass description. The second way is to describe axial-vector mesons using the same parameters as for the vector mesons. We would like to note that the latter opportunity was used in the previous works for the description of the ground states of axial-vector mesons \cite{86,klev,er}. 
We have calculated the mass spectrum, described quark-meson interaction, and considered a number of processes including ground- and excited-state axial-vector mesons.

The masses of ground-state light-quark mesons appeared to be lower than the experimental values, as far as ground states of strange mesons and all excited states are described in satisfactory agreement with the experimental data. A similar situation also takes place in description of masses of scalar mesons \cite{yud}. One of possible explanations of this discrepancy in the ground states description is contribution of four-quark states in scalar mesons \cite{ach}. A similar mechanism is possible for axial-vector mesons.

The masses of the isoscalar mesons differ from those of the isovector mesons. It may be explained by the mixing of light-quark and s-quark states ($f_1(1285)$ and $f_1(1420)$) \cite{f1}, and the same mechanism for the excited states.

As far as we have used the same parameters as for the vector mesons, interaction of axial-vector mesons with quarks can be successfully described by using the same mixing angles. The calculations presented in the previous section support this statement. The results obtained allow one to describe a set of $\tau$-decays and other processes including axial-vector mesons. Indeed, the calculations for the branching ratio of the process $\tau\rightarrow f_1 \pi \nu_\tau$ presented recently in \cite{Vishneva:2014lla}  lead to good agreement with the experimental values.

Let us note that the model of NJL type with non-local four-quark interaction for description of scalar, pseudoscalar, vector and axial-vector mesons was also developed in \cite{and}. Particularly, the authors of this article have obtained $M_{a'_1}=1465\div1850$ MeV.

\section{Appendix: Definition of the coefficients included in the decay amplitudes.}

In order to describe meson decays in the framework of the extended NJL model, we used a set of coefficients $V_{k=2,3,4}^{lmn}$. One can obtain them from the Lagrangians~\eqref{eq:lva},~\eqref{eq:lsp} and they read

\begin{eqnarray}
V_k^{a_1 \pi\rho}&=&\left[\frac{\sin(\chi+\chi_0)}{\sin(2\chi_0)}\right]^2 g_\rho^2 I_k(m_u)\\ \nonumber
&+& 2\frac{\sin(\chi+\chi_0)}{\sin(2\chi_0)}\frac{\sin(\chi-\chi_0)}{\sin(2\chi_0)}g_\rho g_{\rho'}I_k^f(m_u)\\ \nonumber
&+&\left[\frac{\sin(\chi-\chi_0)}{\sin(2\chi_0)}\right]^2 g_{\rho'}^2 I_k^{f^2}(m_u);\\
V_k^{a'_1 \pi\rho}&=&V_k^{a_1 \pi \rho'}=V_k^{a'_1 \pi f_1}=V_k^{a_1 \pi f'_1}=\\ \nonumber
&-&\frac{\sin(\chi+\chi_0)}{\sin(2\chi_0)} \frac{\cos(\chi+\chi_0)}{\sin(2\chi_0)} g_{\rho}^2 I_k(m_u)\\ \nonumber
&-& \frac{\sin(\chi-\chi_0)}{\sin(2\chi_0)} \frac{\cos(\chi+\chi_0)}{\sin(2\chi_0)} g_{\rho} g_{\rho'} I_k^f(m_u)\\ \nonumber
&-&  \frac{\sin(\chi+\chi_0)}{\sin(2\chi_0)} \frac{\cos(\chi-\chi_0)}{\sin(2\chi_0)} g_{\rho} g_{\rho'} I_k^f(m_u)\\ \nonumber
&-&   \frac{\sin(\chi-\chi_0)}{\sin(2\chi_0)} \frac{\cos(\chi-\chi_0)}{\sin(2\chi_0)} g_{\rho'}^2 I_k^{f^2}(m_u);\\ \nonumber
\end{eqnarray}

\begin{eqnarray}   
V_k^{a_1 a_1 \rho}&=&\left[\frac{\sin(\chi+\chi_0)}{\sin(2\chi_0)}\right]^3 I_k(m_u)\\ \nonumber
&+&3\left[\frac{\sin(\chi+\chi_0)}{\sin(2\chi_0)}\right]^2\frac{\sin(\chi-\chi_0)}{\sin(2\chi)} I^f_k(m_u)\\ \nonumber
&+&3\left[\frac{\sin(\chi-\chi_0)}{\sin(2\chi_0)}\right]^2\frac{\sin(\chi+\chi_0)}{\sin(2\chi)} I^{f^2}_k(m_u)\\ \nonumber
&+&\left[\frac{\sin(\chi-\chi_0)}{\sin(2\chi_0)}\right]^3 I^{f^3}_k(m_u);\\
V_k^{a_1 a'_1 \rho}&=&-\left[\frac{\sin(\chi+\chi_0)}{\sin(2\chi_0)}\right]^2\frac{\cos(\chi+\chi_0)}{\sin(2\chi_0)} I_k(m_u)\\ \nonumber
&-&\left[\frac{\sin(\chi+\chi_0)}{\sin(2\chi_0)}\right]^2\frac{\cos(\chi-\chi_0)}{\sin(2\chi_0)} I^f_k(m_u)\\ \nonumber
&-&2\frac{\sin(\chi+\chi_0)}{\sin(2\chi_0)}\frac{\sin(\chi-\chi_0)}{\sin(2\chi)}\frac{\cos(\chi+\chi_0)}{\sin(2\chi_0)} I^f_k(m_u)\\ \nonumber
&-&\left[\frac{\sin(\chi-\chi_0)}{\sin(2\chi_0)}\right]^2\frac{\cos(\chi-\chi_0)}{\sin(2\chi_0)} I^{f^2}_k(m_u)\\ \nonumber
&-&2\frac{\sin(\chi+\chi_0)}{\sin(2\chi_0)}\frac{\sin(\chi-\chi_0)}{\sin(2\chi)}\frac{\cos(\chi-\chi_0)}{\sin(2\chi_0)} I^{f^2}_k(m_u)\\ \nonumber
&-&\left[\frac{\sin(\chi-\chi_0)}{\sin(2\chi_0)}\right]^2\frac{\cos(\chi-\chi_0)}{\sin(2\chi_0)} I^{f^3}_k(m_u);\\ \nonumber
\end{eqnarray}
\begin{equation}
V_k^{a_1\pi}=\frac{\sin(\chi+\chi_0)}{\sin(2\chi_0)} g_\rho I_k(m_u)+\frac{\sin(\chi-\chi_0)}{\sin(2\chi_0)} g_{\rho'} I^f_k(m_u);
\end{equation}
\begin{eqnarray}
V^k_{a_1\pi\gamma}&=&g_\pi\left[ g_\rho\frac{\sin(\chi+\chi_0)}{\sin(2\chi_0)}I_4(m_u)\right.\\ \nonumber
&+&\left.g_{\rho'}\frac{\sin(\chi-\chi_0)}{\sin(2\chi_0)}I_4^f(m_u) \right];\\ 
V^k_{a'_1\pi\gamma}&=&-g_\pi\left[ g_\rho\frac{\cos(\chi+\chi_0)}{\sin(2\chi_0)}I_4(m_u)\right.\\ \nonumber
&-&\left.g_{\rho'}\frac{\cos(\chi-\chi_0)}{\sin(2\chi_0)}I_4^f(m_u) \right];\\ \nonumber
\end{eqnarray}
\begin{eqnarray}
V_k^{a_1{\pi}'\gamma}&=&-\frac{\sin(\chi+\chi_0)}{\sin(2\chi_0)} \frac{\cos(\alpha+\alpha_0)}{\sin(2\alpha_0)} g_\pi g_\rho I_k(m_u)\\ \nonumber
&-&\frac{\sin(\chi-\chi_0)}{\sin(2\chi_0)} \frac{\cos(\alpha+\alpha_0)}{\sin(2\alpha_0)} g_\pi g_{\rho'} I_k^f(m_u)\\ \nonumber
&-&\frac{\sin(\chi+\chi_0)}{\sin(2\chi_0)} \frac{\cos(\alpha-\alpha_0)}{\sin(2\alpha_0)} g_{\pi'} g_\rho I_k^f(m_u)\\ \nonumber
&-&\frac{\sin(\chi-\chi_0)}{\sin(2\chi_0)} \frac{\cos(\alpha-\alpha_0)}{\sin(2\alpha_0)} g_{\pi'} g_{\rho'} I_k^{f^2}(m_u);\\ 
V_k^{a'_1{\pi}'\gamma}&=&\frac{\cos(\chi+\chi_0)}{\sin(2\chi_0)} \frac{\cos(\alpha+\alpha_0)}{\sin(2\alpha_0)} g_\pi g_\rho I_k(m_u)\\ \nonumber
&+&\frac{\cos(\chi-\chi_0)}{\sin(2\chi_0)} \frac{\cos(\alpha+\alpha_0)}{\sin(2\alpha_0)} g_\pi g_{\rho'} I_k^f(m_u)\\ \nonumber
&+&\frac{\cos(\chi+\chi_0)}{\sin(2\chi_0)} \frac{\cos(\alpha-\alpha_0)}{\sin(2\alpha_0)} g_{\pi'} g_\rho I_k^f(m_u)\\ \nonumber
&+&\frac{\cos(\chi-\chi_0)}{\sin(2\chi_0)} \frac{\cos(\alpha-\alpha_0)}{\sin(2\alpha_0)} g_{\pi'} g_{\rho'} I_k^{f^2}(m_u), \nonumber
\end{eqnarray}

\begin{eqnarray}
V^{f_1 a_0 \pi}_k&=&\frac{\sin(\chi+\chi_0)}{\sin(2\chi)} \frac{\sin(\alpha+\alpha_0)}{\sin(2\alpha_0)} g_\sigma g_\rho I_k(m_u)\\ \nonumber
&+&\frac{\sin(\chi-\chi_0)}{\sin(2\chi)} \frac{\sin(\alpha+\alpha_0)}{\sin(2\alpha_0)} g_\sigma g_{\rho'} I_k^f(m_u)\\ \nonumber
&+&\frac{\sin(\chi+\chi_0)}{\sin(2\chi)} \frac{\sin(\alpha-\alpha_0)}{\sin(2\alpha_0)} g_{\sigma'} g_\rho I_k^f(m_u)\\ \nonumber
&+&\frac{\sin(\chi-\chi_0)}{\sin(2\chi)} \frac{\sin(\alpha-\alpha_0)}{\sin(2\alpha_0)} g_{\sigma'} g_{\rho'} I_k^{f^2}(m_u)
\end{eqnarray}
where $\alpha=59.38^\circ$ and $\alpha_0=59.06^\circ$ are the pion mixing angles. They can be obtained in the same way as the angles for $\rho$ and $a_1$ mesons (see \cite{yaf}). We would like to note that we do not take into account the mixing of the ground and excited states of pions, in the processes involving only ground-state pions, because of its negligible contribution.

\section{Acknowledgements}
We are grateful for A. B. Arbuzov, D. G. Kostunin and \fbox{E. A. Kuraev} for useful discussions.

\end{document}